\documentclass[aps,preprintnumbers,prl,twocolumn,superscriptaddress]{revtex4}

\usepackage{epsfig,latexsym,cancel,amssymb,amsmath}
\usepackage{graphicx}
\usepackage{epstopdf}
\usepackage{hyperref, graphicx, slashed, xcolor, amsmath}

\allowdisplaybreaks

\definecolor{red}{rgb}{1,0,0}

\newcommand{\beq}{\begin{equation}}
\newcommand{\eeq}{\end{equation}}
\newcommand{\bea}{\begin{eqnarray}}
\newcommand{\eea}{\end{eqnarray}}
\newcommand{\nn}{\nonumber\\}

\begin{document}

%\baselineskip=18pt \pagestyle{plain} \setcounter{page}{1}

%\preprint{
%\begin{flushright}
%CALT-TH-2018-027
%\end{flushright}
%}

%\vspace*{1.5cm}

\title{KeV Scale  Frozen-in Self-Interacting Fermionic Dark Matter}

\author{Haipeng An}
\affiliation{Department of Physics, Tsinghua University, Beijing 100084, China}
\affiliation{Perimeter Institute for Theoretical Physics, Waterloo, ON, N2L 2Y5, Canada }
\author{Ran Huo}
\affiliation{Department of Physics and Astronomy, University of California, Riverside, California 92521, USA}
\author{Wanqiang Liu}
\affiliation{Department of Physics, Tsinghua University, Beijing 100084, China}

\begin{abstract}

We present a model in which the dark matter particle is frozen-in at MeV scale. In this model the mediator between the standard model sector and the dark sector can automatically provide a self-interaction for dark matter. The interaction strength is naturally to be the in the region in favor of the cluster mass deficit anomaly. Due to the self-scattering the Lyman-$\alpha$ constraint can be relaxed to $m_D \gtrsim 2 $ keV. In this region the self-interaction and the Fermi pressure both play roles on forming a dark matter core at the center of the dwarf galaxies.

%A few hundred eV fermionic dark matter can potentially solve the core-cusp problem in dwarf galaxies. However, such light dark matter particles, if thermally produced, suffer from the constraints of the Lyman-$\alpha$ forests observation. Here we study a model in which the dark matter relic density is produced through thermal freeze-in process, and then due to the self-interaction in the dark sector, the dark matter particles replicate themselves through the $2\rightarrow4$ processes such that the dark sector becomes significantly colder than the Standard Model sector. The self-scattering processes in the early universe in this model also significantly shorten the free-streaming length of the dark matter particles. We show that the parameter space in favor of solving the core-cusp problem is just in the right region to solve the cluster mass-deficit problem and to avoid the stellar constrants.

\end{abstract}
%\preprint{CALT-TH-2018-027}
\maketitle

{\bf Introduction.}
Cold dark matter (CDM) paradigm, although has been extremely successful in explaining the large scale structure of our universe, is challenged by small scale anomalies from dwarf galaxies to galactic clusters. In particular, simulations based on CDM show that the mass density profile for CDM halo increases as $\rho_{\rm DM} \propto r^{-1}$ toward the center~\cite{Dubinski:1991bm,Navarro:1995iw,Navarro:1996gj}, whereas many observed rotation curves of disk galaxies prefer a constant cored density profile $\rho_{\rm DM} \propto r^0$~\cite{Flores:1994gz,Moore:1994yx,Moore:1999gc}. CDM also predicts a greater number of galactic satellites than predicted~\cite{Moore:1999nt}. As for galactic clusters observations show that there is a mass deficit in the inner ${\cal O}$(10) kpc region~\cite{Newman:2012nw} compared to the NFW profile. In \cite{2013MNRAS.429L} the multi-tracer technique was used and the size of the core of Fornax is determine to 2$\sigma$ level to be $0.2~{\rm kpc} < r_c < 2.6$ kpc. However, recent study in \cite{2018MNRAS.429L} shows that the multi-tracer technique can mis-identify a cuspy profile to a cored profile. A more recent study of the dynamical friction of the globular clusters in the Fornax system gives an upper bound of the core size $r_c < 282$ pc~\cite{Boldrini:2018uja}, which favors a small core.

%These small scale anomalies might be understood within the framework of CDM by taking into account baryonic processes, i.e. gas cooling, star formation, and  supernovae feedback in simulations. However, it remains unclear whether the baryonic effects can solve the small scale anomalies, due to the difficulty of modeling the baryonic feedback properly in simulations.

Dark matter (DM), although supported by various evidences from astrophysics to cosmology, appears only through gravitational effects. The small scale properties may provide us with opportunities to explore the particle physics natures of DM.
%Possible particle physics solutions to the small scale anomalies are warm dark matter~\cite{Dalcanton:2000hn,Bode:2000gq}, self-interaction dark matter~\cite{Spergel:1999mh} (and \cite{Tulin:2017ara} for a review), or boson degeneracy~\cite{Ji:1994xh,Hu:2000ke,Peebles:2000yy}.
An interesting observation is that if DM is composed of ${\cal O}(100)$ eV fermions, the Fermi pressure forces the core of dwarf galaxies to be larger than observed (the Tremaine-Gunn bound)~\cite{Tremaine:1979we}. This observation unavoidably leads people to consider the idea that if the mass of DM is just around the boundary of where the Tremaine-Gunn bound allows, the Fermi pressure may provide a solution to the core-cusp problem~\cite{Hogan:2000bv,Destri:2012yn,Domcke:2014kla,Alexander:2016glq,Randall:2016bqw,Giraud:2018gxl}. On the other hand, the Lyman-$\alpha$ forests observation shows that the mass of DM, $m_D$ has to be larger than about 5 keV if the DM particles are thermally produced. keV scale DM particles can also be copiously produced inside stars, and as a result change the life-time, neutrino flux and luminosities of the stars, which strongly constrain the parameter space of this kind of models.

In this work, we propose a DM model based on the freeze-in mechanism~\cite{Hall:2009bx}. In this model we extend SM with a Dirac fermion $\chi$, the DM candidate with mass $m_D\sim$ keV and a massive vector boson $V$ (the dark photon) with a mass $m_V$ at MeV scale. The interaction between the dark sector and the SM sector is conducted by the kinetic mixing between $V$ and the photon field.
The Lagrangian is
\bea\label{eq:lagrangian}
{\cal L} &=& \bar\chi i \gamma^\mu(\partial_\mu - i e_D V_\mu) \chi - m_D \bar\chi\chi \nn
&&- \frac{1}{4} V^{\mu\nu} V_{\mu\nu} + \frac{1}{2}m^2_V V^\mu V_\mu - \frac{1}{2}\kappa V_{\mu\nu} F^{\mu\nu}  \ .
\eea
We show that in this model the Fermi pressure together with the self-scattering can produce a small core in dwarf galaxies, while the self-scattering is naturally in the region in favor of the cluster mass deficit anomaly. We also show that in this model right after frozen-in the DM particles quickly replicate themselves induces a much lower temperature in the dark sector than in the SM sector. The self-scattering of $\chi$ turn free-streaming into Brownian motion, which shortens the distance the DM particles migrate. With these effects the constraint from Lyman-$\alpha$ forests observation can be relaxed.
We also show that the stellar constraints can also be avoided in this model.

Other possible particle physics scenarios in solving the small scale anomalies are warm dark matter~\cite{Dalcanton:2000hn,Bode:2000gq}, self-interaction dark matter~\cite{Spergel:1999mh,Kaplinghat:2015aga} (and \cite{Tulin:2017ara} for a review), or boson degeneracy~\cite{Ji:1994xh,Hu:2000ke,Peebles:2000yy}. The DM models which can solve the anomalies at different scales are self-interaction model with light mediator~\cite{Kaplinghat:2015aga}, the self-scattering through $s$-channel resonance~\cite{Chu:2018fzy}.

{\bf Freeze-in. }
\label{sec:production}
The most important freeze-in channels are the $e^+ e^-$ annihilation channel and the plasmon decay channels.
The production rate of the dark matter number density $n_\chi$ in the $e^+ e^-$ annihilation channel can be written as
%\bea
%\Gamma^\chi_{e^\pm} = \prod_{i=e^\pm\chi\bar\chi}\int \frac{d^3 p_i}{(2\pi)^3 2p_i^0} f_{e^+}(p_{e^+}^0) f_{e^-}(p_{e^{-}}^0) (2\pi)^4 \delta^4(p_{e^+}+p_{e^-} - p_{\chi} - p_{\bar\chi}) \sum_{\rm spins} |{\cal M}_{e^+e^{-}}|^2 \ , \nn
%\eea
%where ${\cal M}$ is the matrix element and $f_{e^+} = f_{e^-} = (e^{E/T}+1)^{-1}$ is the Fermi-Dirac distribution. In this work we are interested in ${\cal O}(100)$ eV scale dark matter particles, and the $e^+ e^-$ annihilation process stops at the temperature smaller than MeV scale. Therefore we can neglect $m_D$ in this calculation. After straightforward calculation we can get
\bea\label{eq:Gammaepem}
\frac{d\Gamma^\chi_{e^\pm}}{d\omega} \!= \!\frac{2\kappa^2 \alpha\alpha_D}{3\pi^2}\!\!\int \!\!dq  \frac{q^2 (s + 2 m_e^2)s} {(s-m_V)^2 + \frac{1}{9}\alpha_D^2s^2} f\left(\frac{\omega}{T},\frac{q}{T},s\right)\!,
\eea
where $s = {\omega}^2 - q^2$ is the center-of-mass energy of the $e^+ e^-$ pair,
\bea
f(x,y,s) = \frac{1}{2\pi y} \frac{4~ {\tanh}^{-1}\left[ \left(\frac{a-1}{a+1}\right)\tanh\left(\frac{b}{2}\right) \right]}{(a-1)(a+1)} \ ,
\eea
with
$a = e^{x/2}$ and $b = \frac{y}{2} \left( 1 -  \frac{4m_e^2}{s}\right)^{1/2}$. In the case $m_V > 2 m_e$, this process is dominated by the production of on-shell $V$, and the production rate is the same as discussed in \cite{Fradette:2014sza}.

%For $T = 10 m_e$, $\alpha_D = 0.01$, $m_D = 1$ keV, the freeze-in rate $dn_\chi/dt$ from the $e^+e^-$ annihilation channel as a function of $m_V$ is shown as the black curve in Fig.~\ref{fig:dndt}. One can see that for $m_V > 2 m_e$ the on-shell production of $V$ through the annihilation of $e^+ e^-$ dominates, this is in the same case as discussed in \cite{Fradette:2014sza}. Whereas in the region $m_V < 2 m_e$ it becomes significantly smaller due to that it can only go through off-shell $V$.

%\subsection{Plasma resonant decay}

Due to the plasma effect transverse photons develop a non-trivial dispersion relation and therefore can decay into a pair of dark matter particles. The collective motion of the charged particles in the thermal plasma behaves like a longitudinal mode of the photon field, which also decays into a $\chi\bar\chi$ pair.
The production rate of the DM number density can be approximately written as~\cite{Braaten:1993jw}
\bea\label{eq:Gammatl}
\frac{d\Gamma^\chi_t}{d\omega} &=& \frac{\kappa^2 \alpha_D}{3\pi^2} \frac{q Z_t}{e^{\omega / T} -1 } \frac{s^3 }{(s-m_V^2)^2 + \frac{1}{9}\alpha_D^2 s^2} \nn
\frac{d\Gamma^\chi_{\ell}}{d\omega} &=& \frac{\kappa^2\alpha_D}{6\pi^2}   \frac{q Z_\ell}{e^{\omega/T} - 1} \frac{  \omega^2 s^2 }{(s - m_V^2)^2 + \frac{1}{9}\alpha_D^2 s^2} \
\eea
where $Z_{t,\ell}$ are the wave function renormalization factors. The magnitude of the three-momentum of the plasmon $q$ can be calculated from the dispersion relations of the transverse and longitudinal plasmons given in \cite{Braaten:1993jw}.

%The contribution to $dn/dt$ from the transverse and longitudinal plasmon decay processes are shown by the red and blue curves in Fig.~\ref{fig:dndt}. The peak in the red curve shows the region where the four-momentum of the transverse plasmon (photon) matches the four-momentum of $V$. As a result in this region $V$ can be resonantly produced. Similarly, in the blue curve the slop in the small $m_V$ region shows that the longitudinal model of $V$ can be resonantly produced and the production rate is proportional to $m_V^2$.
%%This phenomenon is similar to the case of non-relativistic plasma in which the dispersion relation of the transverse and longitudinal plasmons can be written as
%%\bea\label{eq:disNR}
%%\omega_t^2 = k^2 + \omega_p^2\ ,\;\;\; \omega_\ell = \omega_p \ .
%%\eea
%%One can easily see that only at the equality $m_V = \omega_p$ the transverse modes of $V$ can be resonantly produced, whereas the longitudinal mode can be produced if $m_V < \omega_p$.
%Compared to the $e^+ e^-$ annihilation channel one can see that in the region $m_V < 2 m_e$ the contribution from the plasma resonant decay channels is more important.

\begin{figure}
\centering
\includegraphics[height=0.8in]{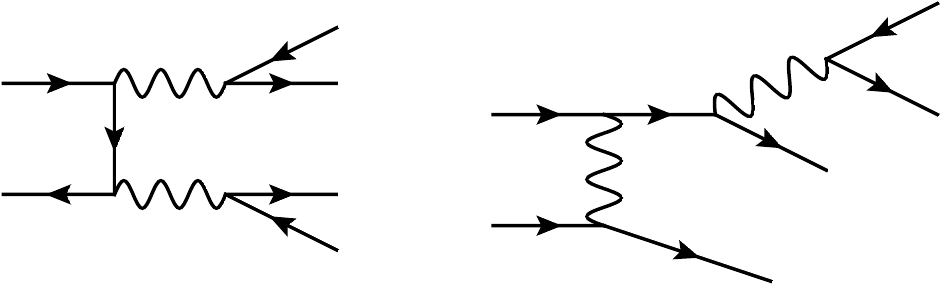}
\caption{Typical Feynman diagrams of 2 to 4 processes.}\label{fig:2to4}
\end{figure}

{\bf The replication.}
%The replication of the DM particles dominated by the $\chi\bar\chi \rightarrow \chi\bar\chi \chi\bar\chi$, $\chi\chi \rightarrow \chi\chi \chi\bar\chi$, and $\bar\chi\bar\chi \rightarrow \bar \chi\bar\chi \bar\chi\chi$ processes.
Right after frozen-in the scattering processes $\chi\bar\chi \rightarrow \chi\bar\chi$, $\chi\chi\rightarrow \chi \chi$ and $\bar\chi\bar\chi \rightarrow \bar\chi\bar\chi$ with a much faster rate than the replication processes establish a ``thermal'' distribution
\bea\label{eq:fD}
f_{\chi(\bar\chi)} = \frac{2}{e^{(E-\mu)/T_D}+1} \ ,
\eea
for $\chi$ and $\bar\chi$ with a chemical potential because the $2\rightarrow2$ processes do not change the numbers of DM and anti-DM particles. In Eq.~(\ref{eq:fD}), $T_D$ is the temperature of the DM particles.
The average kinetic energy of $\chi$ and $\bar\chi$ are around $T_D$. When $T_D$ is larger than $2 m_V$ the replication processes are dominated by the on-shell production of a pair of dark photon $V$, $\chi\bar\chi \rightarrow V V$, with $V$ later decays into a $\chi\bar\chi$ pair. Then when $T_D$ is lowered to be around $m_V$ the $\chi\chi(\bar\chi) \rightarrow \chi\chi(\bar\chi) + V$ process dominates. In the case that $T_D < m_V$, the replication can only go through off-shell dark photons and it is easy to see that cross section in this case goes like $T_D^6/m_V^8$, which diminishes fast with the expansion of the Universe. Therefore the dominant contribution of the replication happens around $T_D \gtrsim m_V$. The typical Feynman diagrams for the 2 to 4 processes are shown in Fig.~\ref{fig:2to4}. In practice we simulate the cross section of the replication processes with CALCHEP~\cite{Belyaev:2012qa}. The replication processes push the dark sector to the true thermal equilibrium with zero chemical potential and with $T_D$ smaller than $T_{\rm SM}$.

\begin{figure}
\centering
\includegraphics[height=1.5in]{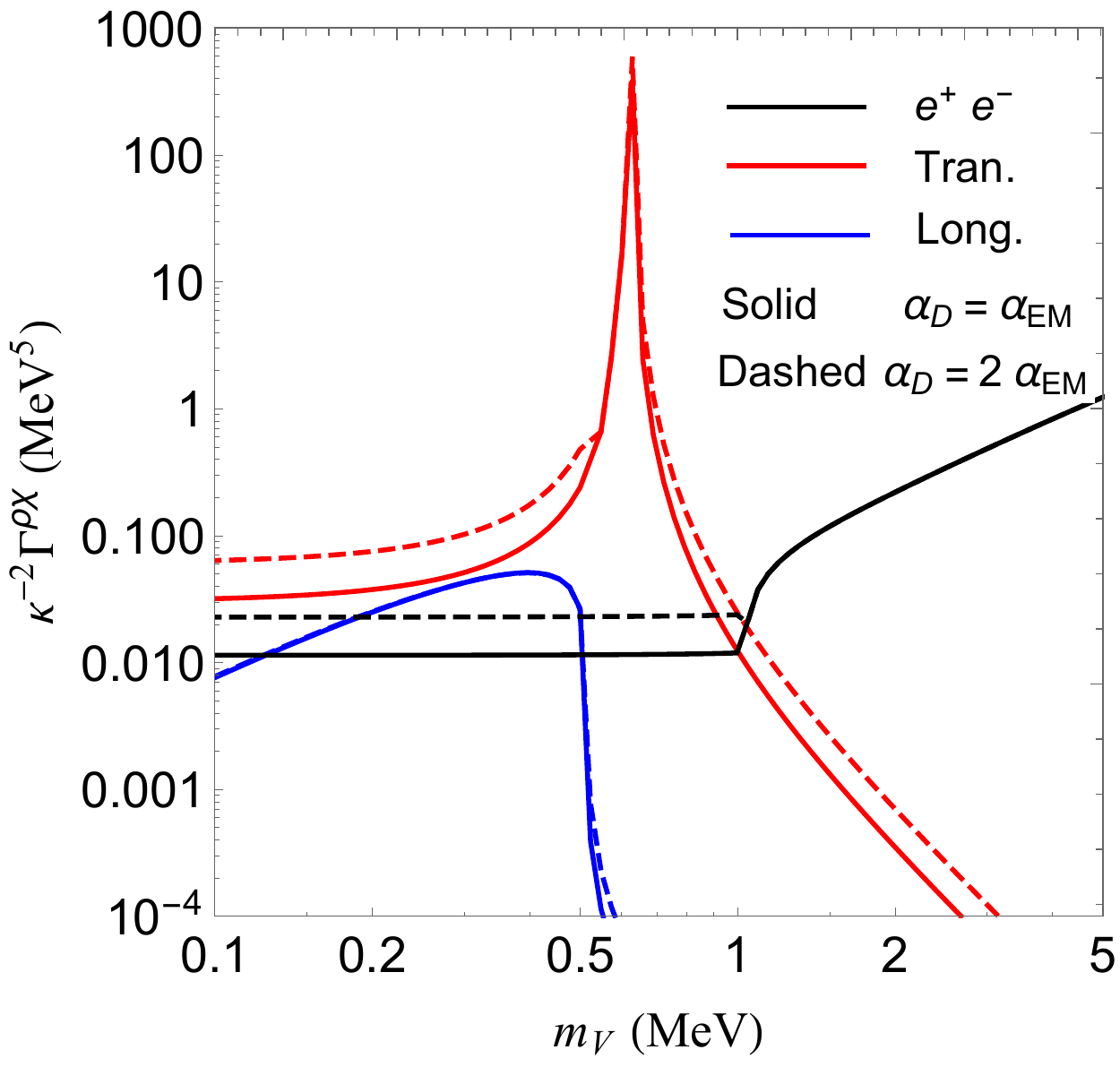}
\includegraphics[height=1.5in]{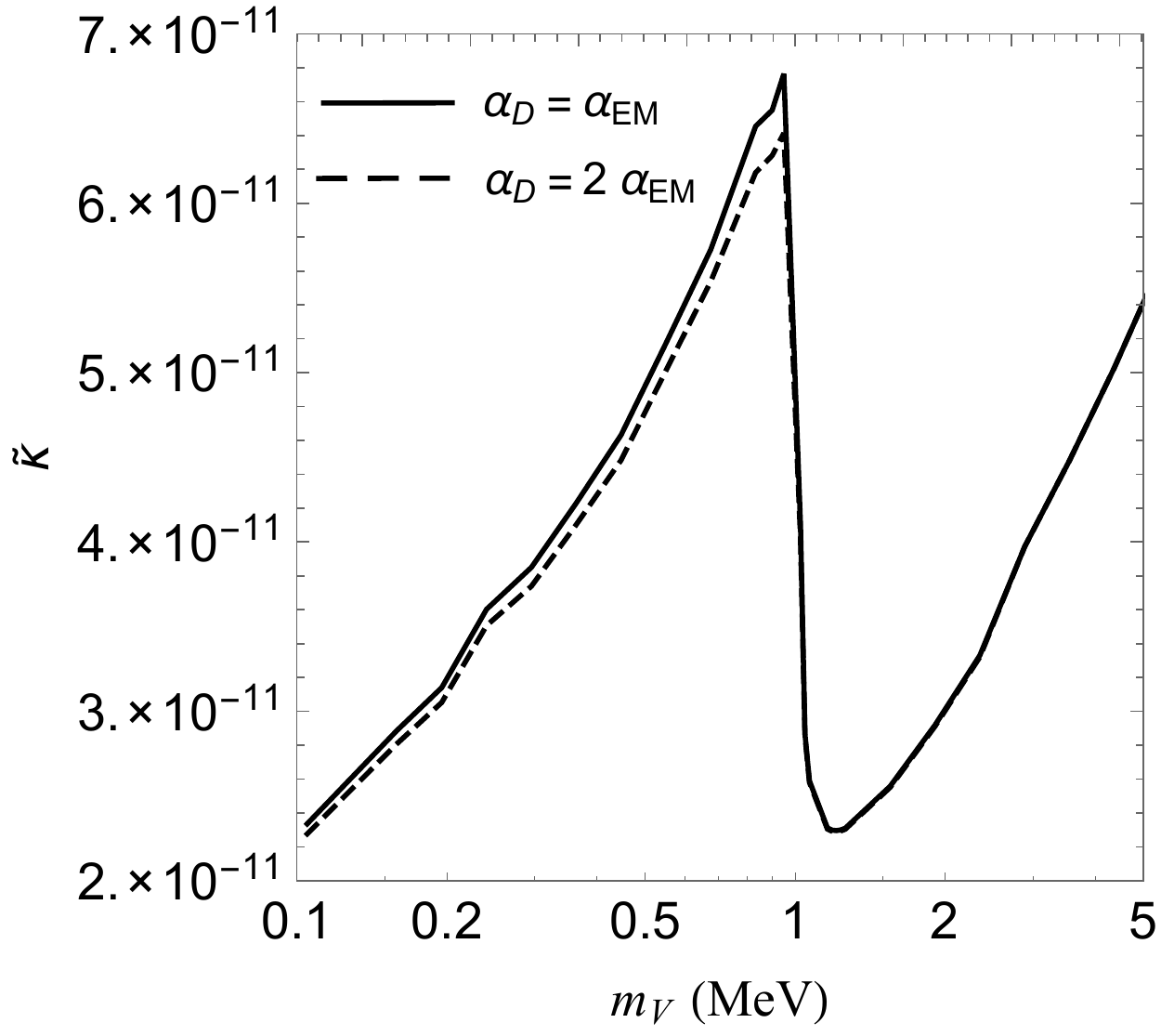}
\caption{Freeze-in processes. Left: freezing-in rate of different processes at $T = 10 m_e$. Right: $\tilde \kappa$ as a function of $m_V$ for $\alpha_D=\alpha_{\rm EM}$ and $\alpha_D=2\alpha_{\rm EM}$.}\label{fig:dndt}
\end{figure}

{\bf The Boltzmann equations.}
\label{sec:boltzmann}
It turns out that in the parameter space in favor of solving the small scale anomalies the $2\rightarrow4$ processes are fast enough that a full thermal equilibrium ($\mu=0$) can be established in the dark sector. The replication processes stop (with a rate smaller than the Hubble expansion rate) when the DM particles are still relativistic. Therefore the DM number density in this case is controlled by the total energy density transferred to the dark sector, one don't need to trace the details of the $2\rightarrow4$ redistribution processes. The Boltzmann equation for freezing-in the DM energy density can be written as
\bea
\frac{d\rho_\chi}{d t} + 4H \rho_\chi = \Gamma^{\rho_\chi}_{e^{\pm}} + \Gamma^{\rho_\chi}_t + \Gamma^{\rho_\chi}_\ell \ ,
\eea
where $\Gamma^{\rho\chi}_{i}$ $(i=e^{\pm},t,\ell)$ can be calculated from Eqs.~(\ref{eq:Gammaepem}) and (\ref{eq:Gammatl}), and their numerical values at $T= 10~m_e$ as functions of $m_V$ are shown in the left panel of Fig.~\ref{fig:dndt}. We can see that at in the case $m_V > 1$ MeV the $e^+ e^-$ annihilation process dominates since $V$ can be produced on-shell. Once $m_V$ is smaller than $1$ MeV, the transverse photon decay starts to dominate. At the region that $V$ can be produced on-shell the rate of the freeze-in processes depends little on $\alpha_D$.

%The Boltzmann equations of the evolution of the dark sector can be written as
%\bea\label{eq:boltzman1}
%\frac{d n_\chi}{dt} + 3 H n_\chi = \Gamma^\chi_{e^\pm}+ \Gamma^\chi_t + \Gamma^\chi_\ell +  {\cal C}_{2\rightarrow4} \ ,
%\eea
%\bea\label{eq:boltzmann2}
%\frac{d\rho_\chi}{d t} + 4H \rho_\chi &=& \Gamma^{\rho_\chi}_{e^{\pm}} + \Gamma^{\rho_\chi}_t + \Gamma^{\rho_\chi}_\ell \ ,
%\eea
%where
%\bea
%{\cal C} &=& (e^{2\mu/T_D} - e^{4\mu/T_D})\frac{T_D}{8\pi^2} \int_0^\infty ds ~s^{3/2}{\cal K}_1\left(\frac{s^{1/2}}{T_D}\right) \nn
%&&\times\left[ \sigma_{\chi\bar\chi\rightarrow 2\chi2\bar\chi} (s) +  \sigma_{2\chi\rightarrow 3\chi1\bar\chi} (s)  \right] \ .
%\eea
%Herewhere ${\cal K}_1$ is the second kind of the modified Bessel function with index one.

In the case of fully thermalization the number density $n_\chi$ has a simple relation with $\rho_\chi$ that (e.g. see \cite{Kolb:1990vq})
\bea\label{eq:rhoN}
n_\chi = 6\times2^{1/4}(15/7)^{3/4}\zeta(3)\pi^{-7/2}\rho_\chi^{3/4} \ .
\eea
Then $m_D$ can be fixed by the DM relic abundance that
\bea\label{eq:8}
\frac{2n_\chi m_D}{n_{\rm B} m_{\rm proton}} \approx \frac{\Omega_D}{\Omega_B}\approx 5 \ .
\eea
Then to get the relic abundance the kinetic mixing $\kappa$ satisfies
\bea\label{eq:kappa}
\kappa = \tilde\kappa(\alpha_D, m_V) \times (m_D/(200~{\rm eV}))^{-2/3} \ .
\eea
$\tilde\kappa$ for $\alpha_D = \alpha_{\rm EM}$ and $\alpha_D = 2\alpha_{\rm EM}$ as a function of $m_V$ is shown in the right panel of Fig.~\ref{fig:dndt}, where one can see that the typical value of $\kappa$ to generate the observed relic abundance of DM is about $10^{-11}$ to $10^{-10}$ and the value of $\tilde\kappa$ depends mildly on $\alpha_D$, since $V$ is preferred to be produced on-shell.

{\bf Stellar constraints.}
\label{sec:constraints}
Dark matter particles with ${\cal O}(100)$ eV mass can be copiously produced inside stars. At the center of the horizontal branch (HB) stars and the red giant (RG) stars the plasma frequency $\omega_p$ is at ${\cal O}(10~{\rm keV})$, which is larger than twice of $m_D$. Therefore the dominant production channel of DM particles is the resonant decay of the transverse photons~\cite{An:2013yfc,An:2013yua}. The energy loss rate to the dark sector is
%The matrix element of this process is
%\bea
%{\cal  M}_{T,L} = \kappa e_D \epsilon_{T,L}^\mu \bar v_\chi(k_1) \gamma_\mu u(k_2) \left(\frac{\omega_p^2}{m_V^2}\right) \ ,
%\eea
%where we make the approximation that $\omega_p^2 \ll m_V^2$. In the NR limit the dispersion relations of the transverse and longitudinal modes are $\omega^2 = \vec k^2 + \omega_p^2$ and $\omega = \omega_p$ with $|\vec k| < \omega_p$. Then the energy loss rate to the dark sector can be written as
\bea\label{eq:stellar}
\frac{d\rho_\chi^{T,L}}{d t} \approx \frac{\kappa^2 \alpha_D}{3\pi^2} \frac{\omega_p^9}{m_V^4}
 \left[f\left(\frac{\omega_p}{T}\right) +
\frac{1}{30}\left(\frac{1}{e^{\omega_p/T} - 1}\right)\right]\ ,
\eea
where
\bea
f(x) = \int_1^\infty dx \frac{x(x^2 - 1)^{1/2}}{e^{(\omega_p /T)x}-1} \ .
\eea
At the center of the HB stars $\omega_p \approx 5$ keV, $T \approx 10$ keV and the density $\rho \approx 3\times 10^4 {~\rm gram ~cm^{-3}}$~\cite{Dearborn:1989he}, and the constraint is that the energy loss rate to the dark sector to be smaller than 8 erg$/$gram$/$sec~\cite{Redondo:2013lna}.
%the constraint from HB stars on $\kappa$ as a function of $m_V$ for $\omega_p \gg m_D$ is shown as the blue curves in Fig.~\ref{fig:stellar}.
For the RG stars we use $\omega_p \approx 20$ keV, $T\approx 8.6$ keV, $\rho \approx 10^6$ gram cm$^{-3}$~\cite{Raffelt:1994ry}, and require that the dark radiation rate to be smaller than 10 erg$/$gram$/$sec~\cite{Redondo:2013lna}.
%The upper limit of $\kappa$ as a function of $m_V$ is shown as the red curve in Fig.~\ref{fig:stellar}. The constraint from the RG stars is about two orders of magnitude stronger than from the HB stars. It is because the dark radiation rate is proportional to $(\omega_p/m_V)^4$. $\omega_p$ at the center of the Sun is about 300 eV, as a result the constraint from the Sun is much weaker.
%
\begin{figure}
\centering
\includegraphics[height=2in]{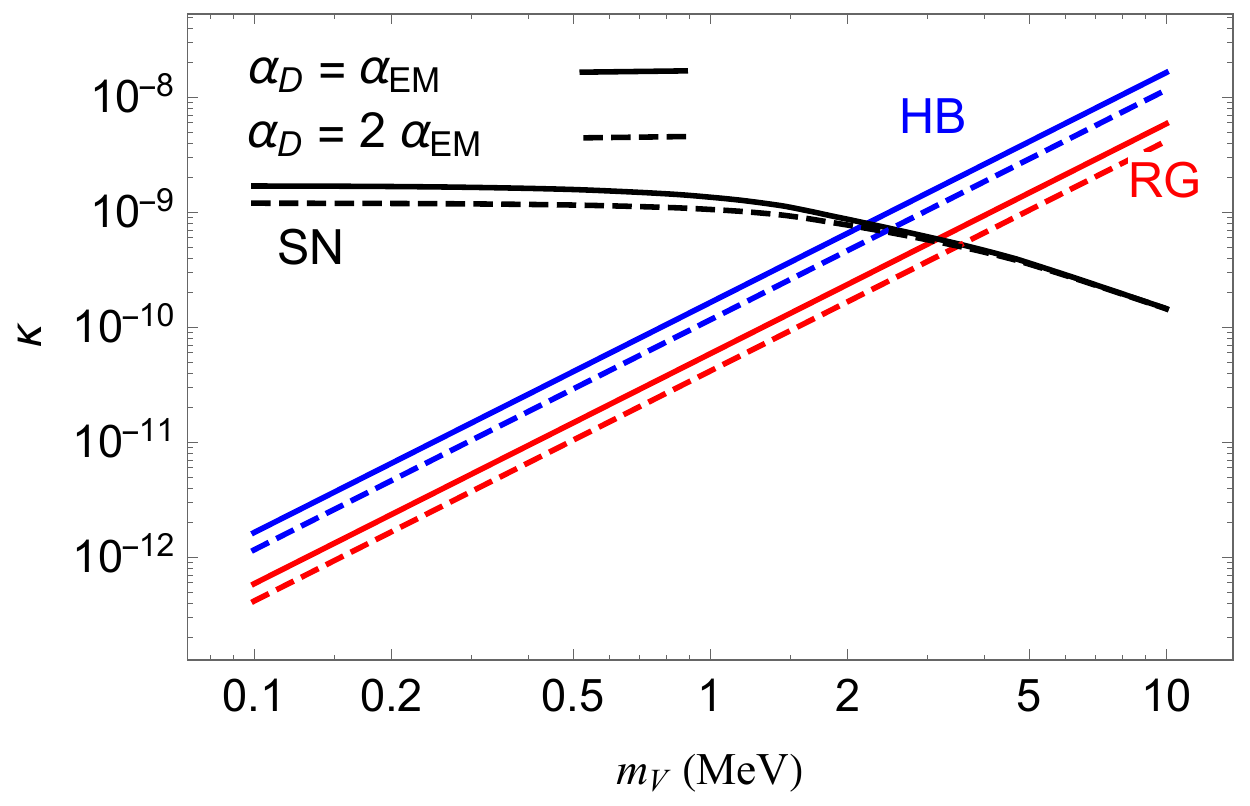}
\caption{Stellar constraints from red giants (red), horizontal branch stars (blue) and supernovae (black).}\label{fig:stellar}
\end{figure}

Supernova1987A (SN) with $\omega_p\sim 10$ MeV and $T\sim 20$ MeV at the center can copiously produce $V$ on-shell if $m_V$ is around $\omega_p$. On the other hand if $m_V \ll \omega_p$ the production rate of the transverse and the longitudinal modes of $V$ are further suppressed by $(m_V/\omega_p)^4$ and $(m_V/\omega_p)^2$, respectively. In this region the dominant dark radiation production channel is again the decay of plasmon into $\chi\bar\chi$ through an off-shell $V$.
We re-interpret the result in \cite{Chang:2016ntp,Chang:2018rso} and its constraint turns out to be weaker than the constraint from RG stars at the region $m_V\sim$ 1 MeV, but it becomes stronger with larger values of $m_V$ as shown in Fig.~\ref{fig:stellar}.

%One can see that the SN constraint is stronger than the constraint from RG stars for $m_V$ larger than a few MeV. This is because in this region $V$ is preferably to be on-shell produced in SN. With smaller $\alpha_D$ the SN constraint will be more important since the direct production of $V$ is independent of $\alpha_D$ whereas the production of $\chi\bar\chi$ through virtual $V$ is proportional to $\alpha_D$ as shown in Eq.~(\ref{eq:stellar}).

{\bf Lyman-alpha forests observation.}
Observations of the absorption lines in the spectra of quasars due to small hydrogen clouds - the so-called Lyman-$\alpha$ forests show the matter power spectrum is not suppressed at Mpc scale. This gives a strong constraint on the free-streaming length of DM. From Ref.~\cite{Irsic:2017ixq} the most aggressive Lyman-$\alpha$ forests bound on the warm DM mass is $m_{\rm WDM} > 5.3$ keV, if an alternative model of the evolution of the inter-galactic matter is taken the constraint can be weaken to $m_{\rm WDM} > 3.5$ keV. In our model the dark sector reaches thermal equilibrium due to the 2 to 4 processes. The constraint on $m_D$ can be weaker than the warm DM. The reason is that the DM particles frequently scatter with each other and their path become a random walk before decoupling, such random walk significantly delay the starting of the free-streaming of the DM particles.

% This also significantly shortens the distance the DM particles migrate before the structure formation.

To obtain an estimation on the parameter space of this model detailed simulation of structure formation is needed, which is beyond the scope of this letter. Here we work on the following simplified treatment. We first use the package CAMB~\cite{Lewis:1999bs} to calculate the matter power spectrum for 5.3 (3.5) keV warm DM model and convert to the 1D matter power spectrum by integration over a $k$ plane. Then similarly we calculate the matter power spectrum for our model, but the free-streaming is only turned on when the temperature is below temperature $T_{\rm fs}$, or the free streaming velocity is simply set to zero when $T>T_{\rm fs}$. $T_{\rm fs}$ is defined as the temperature of the SM sector at which the scattering rate of the DM particles is equal to the Hubble expansion rate. In the NR limit the average cross section of the $\chi\chi$ and $\chi\bar\chi$ processes reads
\bea\label{eq:sigmaS}
\bar\sigma_{\rm nr} = \frac{15\pi \alpha_D^2 m_D^2}{m_V^4} \ .
\eea
Therefore, at temperature $T$ the collision rate can be estimated as
\bea
\Gamma_c(T) = \langle \sigma_{\rm nr} v_r \rangle n_D \approx \frac{30 \zeta(3) \alpha_D^2
\zeta_D T^4 \Omega_D}{\pi m_V^4 \eta_\gamma m_p \Omega_B}  \ ,
\eea
where $\eta_\gamma\approx 6\times 10^{-10}$ is the baryon-to-photon ratio, and
\bea
\zeta_D \equiv \frac{T_D}{T}\approx\left(\frac{10 \eta_\gamma m_p}{3m_D}\right)^{1/3}\!\!\approx0.1\times\left(\frac{m_D}{2{\rm ~keV}}\right)^{-1/3}\!\!.
\eea
Equating $\Gamma_c$ to the Hubble expansion rate $H = 1.66 g_\star^{1/2} T^2/m_{\rm pl}$ where $g_\star$ is the effective degree of freedom and $m_{\rm pl}$ is the Planck mass, we get
\bea
T_{\rm fs} \approx 2{~\rm eV}\times\left( \frac{\alpha_D}{\alpha_{\rm EM}} \right)^{-1}\left(\frac{\zeta_D}{0.1}\right)^{-1/2} \left(\frac{m_V}{1~{\rm MeV}}\right)^2 \ .
\eea

\begin{figure}
\centering
\includegraphics[height=2in]{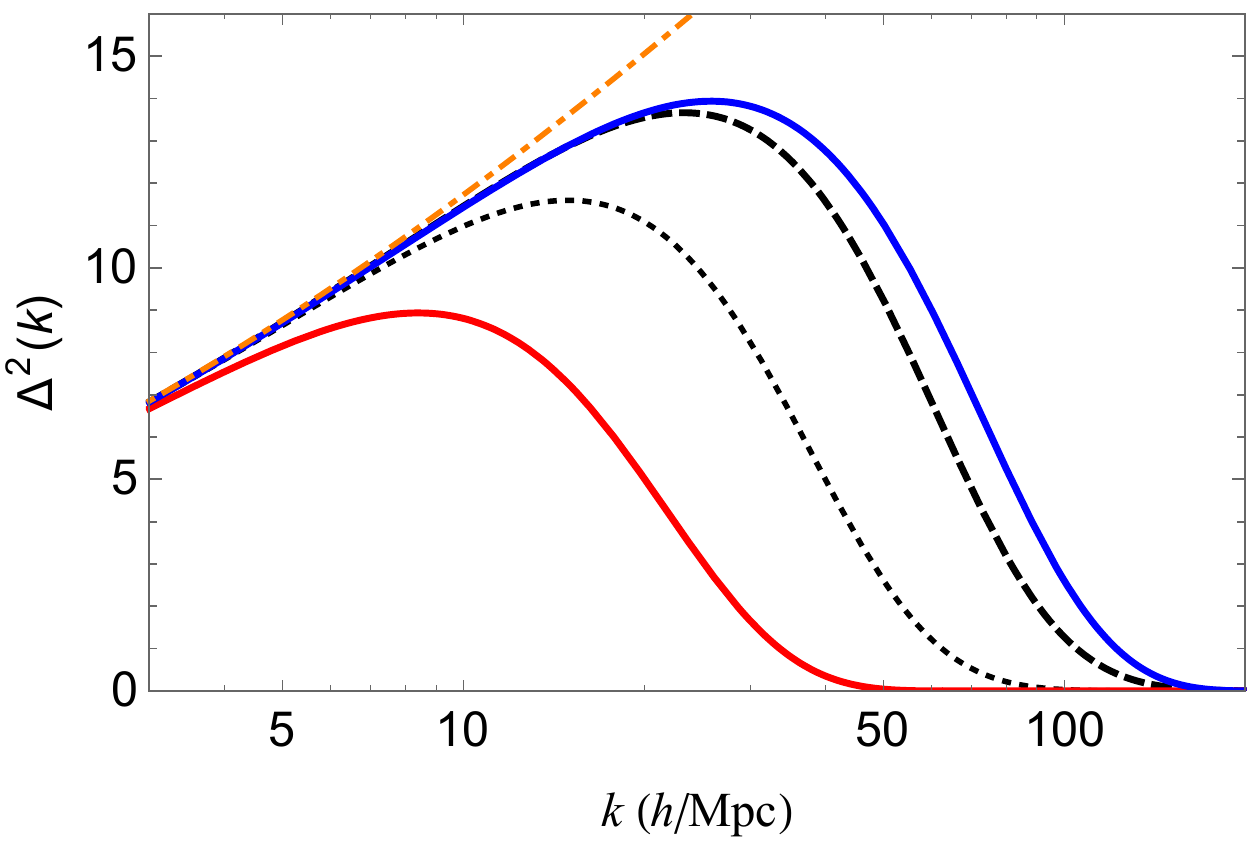}
\caption{1D matter power spectrum simulated by CAMB. The black dashed and dotted curves are for WDM model with $m {\rm WDM} = 5.3$ and 3.5 keV, respectively. The red (blue) curves are for the frozen-in self-interaction DM model with $\sigma_T/m_D$ fixed at 0.1 cm$^2/$gram and $m_D = 1$ keV (3 keV). The orange dot-dashed curve is for the CDM model. }\label{fig:camb}
\end{figure}

The 1D matter power spectrum we simulated using the CAMB package is shown in Fig.~\ref{fig:camb}, where the black dashed and dotted curves are for matter power spectrum with 5.3 keV and 3.5 keV warm DM model. The red and blue curves are for the $m_D=1$ keV and 3 keV in the frozen-in DM model with the self-scattering cross section fixed at $\sigma_T/m_D = 0.1$ cm$^2$/gram chosen in favor of the cluster mass deficit anomaly which will be discussed later.

\begin{figure}
\centering
\includegraphics[height=2in]{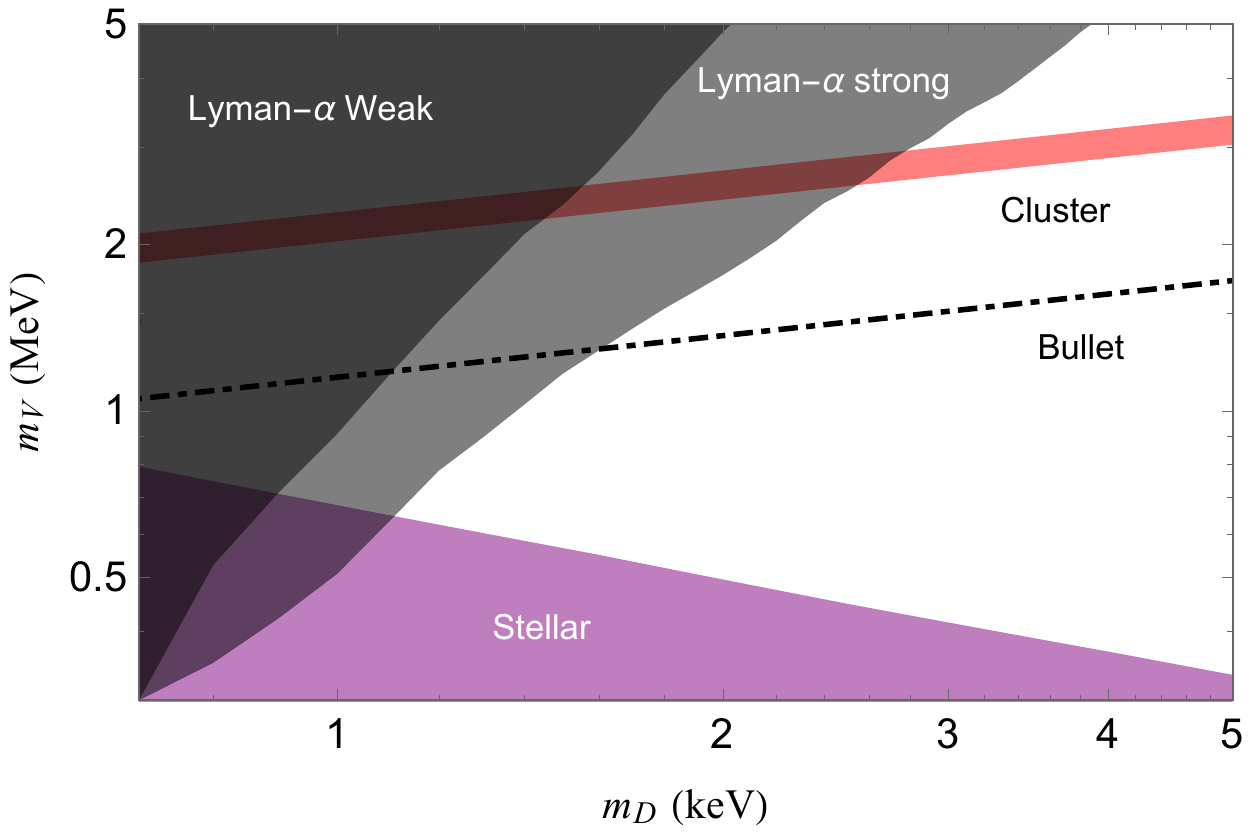}
\caption{Parameter space of the frozen-in self-interacting DM model with $\alpha_D = 2\alpha_{\rm EM}$. The grey regions are excluded by the Lyman-alpha forests observation. The Purple region is excluded by stellar constraints. The kinetic mixing $\kappa$ in this figure is determined by the DM relic abundance. The red band shows the region in favor of the cluster mass deficit problem. The dot-dashed line shows the lower limit from the bullet cluster.}\label{fig:alpha1}
\end{figure}

We scan the parameter space in this model and the result for $\alpha_D = 2\alpha_{\rm EM}$ is shown in Fig.~\ref{fig:alpha1}, where $\kappa$ is determined from the relic abundance. The grey regions show the strong and weak constraints from the Lyman-$\alpha$ observation. The purple region is excluded by the stellar constraints.

%One can see that with the self-scattering and a colder temperature the constraint from Lyman-$\alpha$ forests observation only requires $m_D$ to be larger than about 70 eV. This is well within the region where the Pauli exclusion of the $\chi$ and $\bar\chi$ can potentially solve the core-cusp problem.
%where $\zeta_D$ is from the effect that the dark sector is cooler and the smaller logarithmic factor is due to the self-scattering.

%{\it Bullet cluster bound.}
%The most stringent bound on DM self-interaction is from the bullet cluster requiring $\sigma_T/m_D \lesssim 0.7{~\rm cm^2/gram}$~\cite{Randall:2007ph}. With Eq.~(\ref{eq:sigmaS}) we have in this model
%\bea\label{eq:sigmaT}
%\sigma_T/m_D = 0.1~{\rm cm^2/gram}\times\left( \frac{\alpha_D}{\alpha_{EM}} \right)^2 \times\left(\frac{m_D}{200 {~\rm eV}}\right) \times \left( \frac{m_V}{1~{\rm MeV}} \right)^{-4} \ .
%\eea
%One can see that for typical values of parameters in favor of the core-cusp anomaly of the dwarf spheroidal galaxies the bullet bound is easily satisfied.

\begin{figure}
\centering
\includegraphics[height=2in]{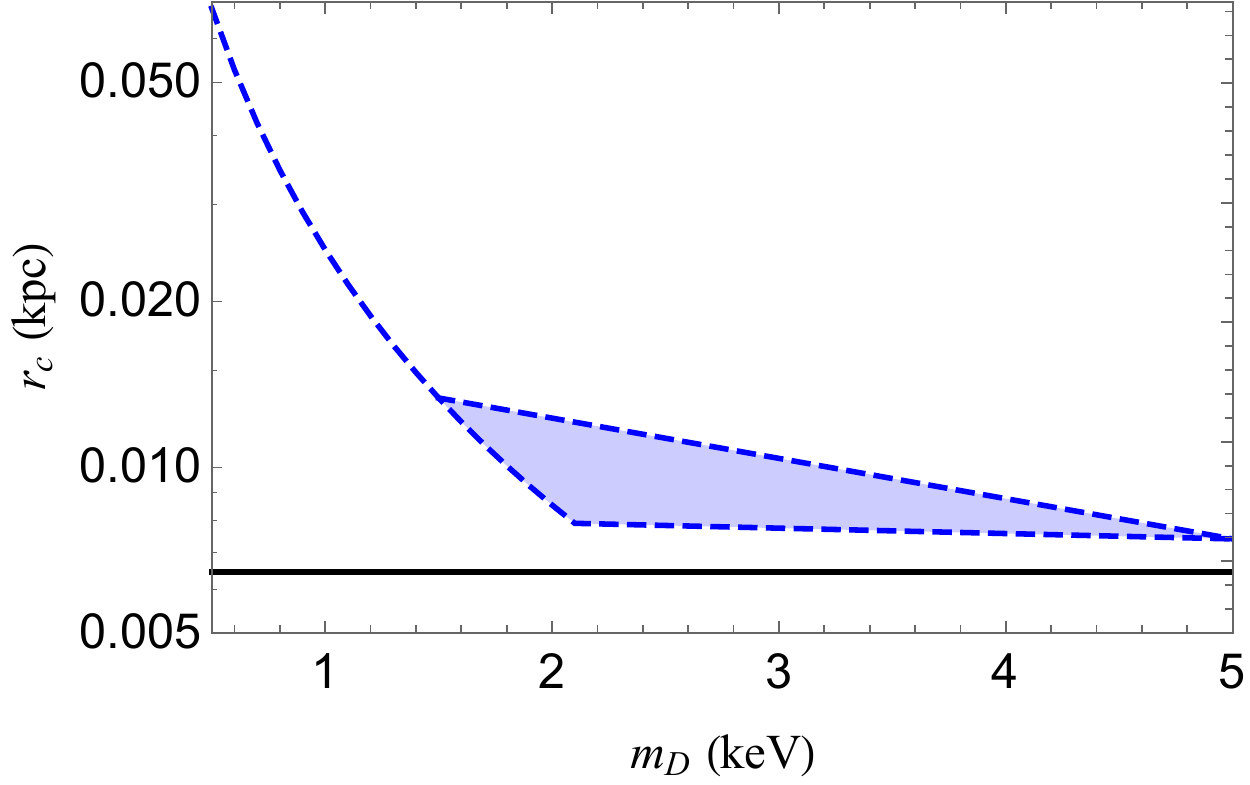}
\caption{Core radius of the Fornax dwarf galaxy as a function of $m_D$ with self-scattering cross section fixed at $0.1$ cm$^2/$gram and the Fermi pressure included. The black line shows the size of the core with only considering the DM self-scattering effect. The data of Fornax is from \cite{Boldrini:2018uja}. }\label{fig:core}
\end{figure}

{\bf The implication on small scale anomalies.}
It has been observed that the profile of clusters agrees well with the NFW process at the region outside ${\cal O}$(10) kpc region from the center, within ${\cal O}$(10) kpc on the other hand a mass deficit is observed. However, it is shown that this problem can be solved if the DM particles have a self-interaction with $\sigma_T/m_D = 0.10^{+0.03}_{-0.02}~{\rm cm}^2/{\rm gram}$~\cite{Kaplinghat:2015aga}, where $\sigma_T$ is the momentum-transfer cross section. But as in \cite{Newman:2012nw}, due to the observation of the out flow and the severe baryonic process at the center of the cluster, observations of cluster alone cannot provide unambiguous support for DM theories. However, in a later study~\cite{Elbert:2016dbb} numerical simulation shows that a self-interaction with $\sigma_T/m_D > 0.1$ cm$^2/$gram is disfavored.
In our model the dark photon $V$ conducts a self-interaction of $\chi$ and $\bar\chi$. Since $m_V \gg m_D$ the scattering is $s$-wave and therefore $\sigma_T$ equals the total cross section. From Eq.~(\ref{eq:sigmaS}),
\bea
\frac{\sigma_T}{m_D}
= 0.125~{\rm cm^2/g}\left(\frac{\alpha_D}{\alpha_{\rm EM}}\right)^2 \left(\frac{m_D}{2{~\rm keV}}\right)\left(\frac{m_V}{2~{\rm MeV}}\right)^{-4},
\eea
which is just in the right region.
The region in favor of the cluster mass deficit problem  for $\alpha_D = 2 \alpha_{\rm EM}$ is shown as the red band in Fig.~\ref{fig:alpha1}. One can see that with considering self-scattering the Lyman-$\alpha$ constraint can be lowered to about 2.5 keV (1.4 keV) for the strong (weak) bound.

DM self scattering cross section will induce a cored profile. As the cross section goes small N body simulation is hard to have resolution to see such core, so here we use the analytical modeling of~\cite{Kamada:2016euw}. In the outer region of a halo the DM scattering count is statistically less than one in the history of the halo, so it will not be significant to change an NFW profile. But in the central region such scattering makes the halo isothermal. In that case in the partition function the momentum part will always gives a constant after integration for Maxwell distribution, so the density $\rho(r)\propto e^{-\frac{\Phi}{\sigma^2}}$. Then Poission equation
\bea\label{eq:Jeans}
\nabla^2\Phi=4\pi G\rho=4\pi G(\rho_0e^{-\frac{\Phi}{\sigma^2}}+\rho_{B})
\eea
can be solved with the observed baryonic distribution. At boundary $r_M$ which is the transition point of the two regions, we impose the physical condition that the enclosed mass as well as the local density are the same as their corresponding NFW values. In fact there are studies shown that to form a thermal distribution at least 2.7 collisions is required~\cite{Monroe:1993zz}. The required number of collisions to reach thermal equilibrium can induce considerable uncertainties in determining the size of the core \footnote{Detailed calculation shows that for $\sigma_T/m_D = 0.1$ cm$^2/$gram, the difference can be as large as one order of magnitude.}. In this study we adopt the criteria of 2.7 collisions. Moreover, in this study the NFW reference halo is taken according to the Planck $\Lambda$CDM halo concentration-mass relation, $c_{200}=10^{0.905\pm0.11}(M_{200}/10^{12}h^{-1}M_\odot)^{-0.101}$~\cite{2014MNRAS.Dutton}. For our fermionic DM model we use Fermi-Dirac distribution instead which also incorporate the Fermi pressure, the difference for a fixed DM mass is here the chemical potential is a new parameter, while the central density $\rho_0$ parameter in the Maxwell case now can be calculated.

For keV-scale fermionic dark matter the Fermi degeneracy pressure will also lead to a sizable core.
In the NR limit the energy density of a Fermi-degenerate gas is $\rho=m_D n= 4\pi g_D m_D p_F^3/3(2\pi)^3$, where $p_F$ is the Fermi momentum and $g_D$ is the degeneracy of DM $(g_D=4)$. Then the pressure can be written as $P=(4\pi g/3(2\pi)^3)\int_0^{p_F}(p^4/m_D)d p = (4\pi^2/5m_D^{8/3})(3/4\pi g_D)^{2/3} \rho^{5/3}$, in terms of $\rho$. The hydrostatic equilibrium equation of the halo gives
\bea\label{eq:hydro}
\frac{d P(r)}{d r} = - \frac{G M(r)}{r^2} \rho(r) \ ,
\eea
where $G$ is the Newton's gravitational constant, $M(r)$ is the mass enclosed within the radius $r$ including the baryonic mass. Eq.~(\ref{eq:hydro}) is an integro-differential equation of $\rho$, which can reduce to the second order differential equation
\bea\label{eq19}
\!-\frac{4}{3}\frac{4\pi^2}{3 m_D^{8/3}}\left(\frac{3}{4\pi g_D}\right)^{2/3}\frac{1}{r^2} \frac{d}{dr}\left( \frac{r^2} {\rho^{1/3}} \frac{d\rho}{dr} \right) \!= \!4\pi G (\rho+\rho_B) \nn
\eea
As such Fermi pressure just affect the inner region of the halo, we should also impose the same boundary condition as above at transition radius, $M(r_M)=M_{\rm NFW}(r_M)$ and $\rho(r_M)=\rho_{\rm NFW}(r_M)$.

In Fig.~\ref{fig:core} we show the radius of the core as a function of $m_D$ with $\sigma_T/m_D$ fixed as 0.1 cm$^2/$gram. The radius of the core, $r_c$, is defined as the radius where the density is half of the density at the center. In getting the plot for small DM mass we solve Eq.~(\ref{eq19}) and for large DM mass we solve Eq.~(\ref{eq:Jeans}), because the former strongly depends on $m_D$ and the core solution gets smaller while using larger $m_D$ whereas the latter one is dominated by scattering effect and is independent of $m_D$. In the shaded region where $m_D$ is between 1.5 to about 5 keV, the classical pressure becomes non-negligible. As a result a sizable correction in this intermediate region of our estimation is expected. The black line on the other hand shows the size of core without consider the effect of the Fermi pressure, namely pure dark matter scattering effect with the Maxwell distribution. One can see that at $m_D=1.5$ keV the size of the core with the Fermi pressure is about 15 parsec, and at $m_D = 2.5$ keV the size of the core can be about 10 parsec.

%A more careful study including the gravitational effect and thermal envelop effect found that suitable dwarf galaxies cores can be achieved with DM mass in the range of 70 eV $-$ 400 eV~\cite{Randall:2016bqw}. This result is based on the assumption that the DM particle is a spin-$1/2$ fermion with two internal degrees of freedom. In our case we have both $\chi$ and $\bar\chi$, therefore our model is equivalent to the $N_f=2$ case in \cite{Randall:2016bqw}. With fixed mass density at the core the radius of the fermi core scales as $(m_D^4 N_f)^{-1/3}$. Therefore in the Dirac DM case the region in favor of solving the core-cusp problem is in the range of 59 eV $-$ 336 eV.

{\bf Conclusion and discussions.}
We have presented a model that the DM relic abundance is generated through the freeze-in mechanism. In this model the Lyman-$\alpha$ constraint can be relaxed to $ m_D \gtrsim$ keV. In this region the Fermi pressure and the self-scattering can produce a small core ($\sim$ 10) pc
in dwarf galaxies.
%With $m_D$ to be around a few hundred eV and $m_V$ to be around 1 MeV, we have shown that the Fermi pressure can be used to solve the core-cusp problem and the self scattering of the DM particles conducted by $V$ is just in the right region to solve the cluster mass deficit problem. In this model the Lyman-$\alpha$ constraint is avoided since the dark sector is colder than the SM sector and the self-scattering of the DM particles shortens the distance the DM particles migrate. Quantitative study shows that stellar constraints are also avoided. Since the self-scattering is $s$-wave its cross section is independent of the DM velocity. Therefore in the region in favor of solving the cluster mass deficit problem the bullet cluster constraint~\cite{Randall:2007ph} can also be avoided.
We use the models in Refs.~\cite{Randall:2016bqw,Kamada:2016euw} to analyze the properties of dwarf galaxies. To get a better understanding a detailed numerical simulation with the Fermi pressure included is needed.

{\bf Acknowledgments}
We thank J. Bramante ,M. Pospelov, M. B. Wise and H.-B. Yu for the useful discussions. This work is supported by  the Recruitment Program for Young Professionals of the 1000 Talented Plan, the Tsinghua University Initiative Scientific Research Program and the National Key Research
and Development Program of China (Grant
No. 2017YFA0402201). Research at Perimeter Institute is supported by the Government of Canada through Industry Canada and by the Province of Ontario through the Ministry of Economic Development \& Innovation.

\end{document}